# Should the research performance of scientists be distinguished by gender?[1]

*Giovanni Abramo[a], Tindaro Cicero[b], Ciriaco Andrea D'Angelo[a,c]*

[a] Laboratory for Studies of Research and Technology Transfer
at the Institute for System Analysis and Computer Science (IASI-CNR)
National Research Council of Italy

[b] Italian National Agency for the Evaluation of Universities and Research Institutes
(ANVUR)

[c] Department of Engineering and Management
University of Rome "Tor Vergata"

**Abstract**

The literature on gender differences in research performance seems to suggest a gap between men and women, where the former outperform the latter. Whether one agrees with the different factors proposed to explain the phenomenon, it is worthwhile to verify if comparing the performance within each gender, rather than without distinction, gives significantly different ranking lists. If there were some structural factor that determined a penalty in performance of female researchers compared to their male peers, then under conditions of equal capacities of men and women, any comparative evaluations of individual performance that fail to account for gender differences would lead to distortion of the judgments in favor of men. In this work we measure the extent of differences in rank between the two methods of comparing performance in each field of the hard sciences: for professors in the Italian university system, we compare the distributions of research performance for men and women and subsequently the ranking lists with and without distinction by gender. The results are of interest for the optimization of efficient selection in formulation of recruitment, career advancement and incentive schemes.

# 1. Introduction

The scientific debate on gender aspects in research systems has focused primarily on the overrepresentation of male academics, often suggesting the occurrence of systematic practices of gender discrimination. The data on the staff of national research systems indeed reveal a significant gap in the presence of women. Only four of 28 OECD nations[2] (OECD 2014) - Portugal, Estonia, Slovak Republic and Iceland – show a percentage of women greater than 40%, and in any case less than 46%. In the UK, women represent 38.3% of total researchers, in Italy 34.5%; in France the share drops below 26%, and in Germany does not reach 25%. In Japan women represent only 13.8% of national research staff. Although the four Nordic countries (Denmark, Finland, Norway, Sweden) are considered as progressive in women's rights, the fact in these nations is that for each female scientist there are two male colleagues.

Alongside the studies illustrating the underrepresentation of women in science there is a major stream of literature that demonstrates the presence of a so-called "productivity gap" in favor of men. The lesser productivity of female researchers has been established in tens of studies of diverse disciplines and countries (Larivière et al., 2013; Mauleón and Bordons, 2006; Xie and Shauman, 2004; Long, 1992; Fox, 1983). Examining the issue in detail, it emerges that gender differences lessen over time (Frietsch et al., 2009; Abramo et al., 2009a; Alonso-Arroyo et al., 2007; Leahey, 2006; Xie and Shauman, 1998; Cole and Zuckerman, 1984) and seem to be most visible in the early career stages (Xie and Shauman, 1998). The tails of the distribution of scientific performance are especially affected by gender differences. The concentration of women among very low performers is greater than that of men (Alonso-Arroyo et al., 2007; Lemoine, 1992), while their representation among top scientists is lower (Abramo et al., 2009b; Bordons et al., 2003). In the area of patenting, women faculty members patent at about 40% of the rate of men (Ding et al., 2006).

However there are a significant number of scientific sectors where the performance of women does not result as inferior (Abramo et al., 2009b). Yet even in these cases, men still predominate in the prestigious first and last author positions of the byline, and women are significantly underrepresented as authors of single-authored papers (West et al., 2013).

Many scholars have inquired into the possible causes of the productivity gap. In general, a researcher's performance depends on his or her capacities, but also derives from a series of gender-dependent environmental and personal factors (Zainab, 1999). Discrimination can emerge in the early stage of the relationships between professors and their students. Moss-Racusin et al. (2012) show the subtle bias in favor of male students that occurs in science faculties. Among the factors that can produce the gender gap, Rossiter (1993) indicated the "Matilda effect",[3] where female scientists active in research are not recognized in the publication bylines. In the career stage of selecting university professors the percentages of female applicants who are successful is generally lower (van der Brink et al., 2006). In the phase of entry to the academic professional environment females generally evaluate their mentors as less satisfactory than do their male colleagues (Sambunjak et al., 2006).

However it is also clear that there are changes in the personal and working contexts of individuals, and that these influence their productivity over time. In the late postdoctoral and early faculty years many qualified women scientists stop applying for NIH grants (Ley and Hamilton, 2008). During their careers, women also present lower productivity in the intermediate levels of seniority (Mauleón et al., 2008). In this stage, differing forms of marriage conduct (Fox, 2005) and the presence of school-age children seem to have a negative effect on research productivity (Fox, 2005; Stacks, 2004; Kyvik and Teigen, 1996). The level of specialization also has a positive relation with research productivity, which could explain a part of the negative gap for women, who are generally less specialized than their male colleagues (Leahey, 2006). It has been verified that research collaborations have a positive correlation with scientific performance (Abramo et al., 2009c; Lee and Bozeman, 2005; Dundar and Lewis, 1998), particularly collaborations at the international level (Barjak and Robinson, 2007; Martin-Sempere et al., 2002; Van Raan, 1998). However female researchers register less international collaborations than men (Abramo et al., 2013a), probably due in part to motivations against travelling in consideration of family roles. In general, women tend to have more restricted collaboration networks than men (Badar et al., 2013; Larivière et al., 2011; Kyvik and Teigen, 1996), particularly in the first years of their career (McDowell et al., 2006; McDowell and Smith, 1992). This limits their access to resources and other complementary assets, necessary for their research activities. In fact academic institutions often do not provide adequate financial support for their female researchers, particularly in the hard sciences (Duch et al., 2012). According to Ceci and Williams (2011) differential gender outcomes result exclusively from differences in resources. When contrasting research performance by gender, one should account for compulsory abstention from work, such as maternity or sick leaves. For large-scale studies investigators often lack such information, which causes a distortion in favor of men.

---

[2] Data for the remaining 6 OECD nations (Australia, Canada, Israel, Mexico, New Zealand, and United States) are not available.

[3] Named for the 19th century social activist, Matilda Joslyn.

However the aim of the current paper is not the further investigation of if or to what extent there is gender discrimination in the research sphere, or to further examine the objective limitations on women's careers given their roles in nuclear families. Instead, our specific objective is to verify if separating the measurement of research performance by gender produces notably different results compared to measurement without such distinction. A female researcher who results less productive than a male when evaluation does not distinguish by gender, may indeed result relatively more productive when research assessment is separated by gender. We then leave it to the decision-maker to choose which approach to adopt, according to the evaluation objectives and the conditions of the context. In those contexts where gender discrimination is understood to exist, or where the family roles of women condition the time, energies and concentration devoted to research, then the conduct of evaluations without distinction by gender would inevitably penalize women. The results of the analysis are of interest for all processes involving efficient selection, such as the formulation of incentive systems in research organizations; methods of evaluation for applicants in career recruitment and advancement, or calls for project proposals.

The context for the study is Italy's national staff of professors in the disciplines of the hard sciences, considered the most appropriate fields for the use of bibliometric techniques in performance evaluation. The Italian context is particularly suitable for the analyses because of its national classification system for faculty members, in which each professor is identified as belonging to one and only one field of research. This feature permits minimization of distortion in the comparative evaluation of researchers working in different research fields, which arises due to the differing intensity of publication across fields, and also permits observation of fluctuations in the variables of interest across the fields. We thus draw up two ranking lists of individual performance for the 2006-2010 period of production in each research field: one with distinction by gender and one without. In a future study we will extend the analyses to the comparison of performance ranking lists at the aggregated level of the university.

The next section provides an overview of the gender profile of all Italian academic staff in the year 2011, the reference year for our analyses. Section 3 presents the methodology adopted for the calculations of productivity and the dataset used for the analyses. Section 4 presents the principle results of the study. The final section provides the conclusions.

**2. The presence of women in the Italian academic staff**

The Italian Ministry of Education, Universities and Research (MIUR) recognizes a total of 95 universities as having the authority to issue legally-recognized degrees. Twenty-nine of these are private, small-sized, special-focus universities. Sixty-seven are public and generally multi-disciplinary universities, scattered throughout the nation. The overall staff system consists of 58,224 professors, of which 94.9% are employed in public universities. As noted above, the Italian higher education system seems unique for a system in which each professor is officially classified as belonging to a single research field. These formally defined fields are called Scientific Disciplinary Sectors (SDSs), of which there are 370 in all, and are grouped into University Disciplinary Areas (UDAs)[4], 14 in all.

According to a 2011 study by the MIUR[5], the Italian university system features a majority of women among both students (ISCED classification 5)[6] and graduates (ISCED 6). However if we observe the makeup of faculty members we see that the relationship is inverted. After completing their education, women more rarely enter an academic career and still more rarely reach higher positions in the academic hierarchy[7]. The data indicate a trend towards remediation: between 1998 and 2010 the presence of women in Italian academic staff increased significantly, although with substantial differences across the disciplines, with the maximum increase in Chemistry and the minimum in Mathematics (CNVSU, 2011). Female professors are in the majority only in the UDAs of Ancient history, philology, literature, art history (55.2%) and Biology (51.6%) (Table 1). The UDAs with the lowest presence of women are Physics (19.6%) and Industrial and information engineering (15.1%).

Gender differences are not homogeneous across the universities. In History, philosophy, pedagogy and psychology, one university shows the situation of 87.5% female staff. On the other hand, there are two UDAs (Industrial and information engineering; Political and social sciences) where some universities have an entirely male staff. An overview of the entire distribution of data (Table 1) reveals notable variance per UDA: the area of History, philosophy, pedagogy, psychology registers the greatest standard deviation (13.0) and shows two extreme cases of universities with presences of women at 13.6% and 87.5%. It is precisely the two UDAs with the lowest female incidence (Physics; Industrial and Information Engineering) that also show the lowest variability in this incidence. An analysis by academic rank reveals a striking trend towards the decrease of female underrepresentation in universities: the incidence of female assistant professors (45.3%) is now much higher than that of full professors (20.7%) (Table 2). In fact in four UDAs (Chemistry; Biology; Ancient history, philology, literature and art history; History, philosophy, pedagogy and psychology) female assistant professors are now more numerous than males.

---

[4] The complete list is available at http://attiministeriali.miur.it/UserFiles/115.htm, last accessed 22/10/2014.
[5] http://statistica.miur.it/scripts/IU/vIU0.asp, last accessed 22/10/2014
[6] International Standard Classification of Education
[7] This phenomenon has also been observed in Netherlands universities (Van der Brink et al., 2006).

*Table 1: Incidence of female professors in the Italian academic system, per UDA; data observed at 31/12/2011, as elaborated from MIUR database*
*http://cercauniversita.cineca.it/php5/docenti/cerca.php, last accessed 22/10/2014*

| UDA* | SDS** | Professors | Female (%) | Incidence of female professors in universities (%) | | | | |
|---|---|---|---|---|---|---|---|---|
| | | | | Universities† | Median | Min | Max | St. dev. |
| 01 | 10 | 3,235 | 33.8 | 50 | 33.3 | 7.7 | 60.0 | 10.8 |
| 02 | 8 | 2,288 | 19.6 | 44 | 18.1 | 5.6 | 34.7 | 7.5 |
| 03 | 12 | 2,941 | 43.7 | 43 | 43.8 | 20.0 | 63.6 | 10.2 |
| 04 | 12 | 1,086 | 28.7 | 30 | 27.4 | 11.1 | 62.5 | 11.0 |
| 05 | 19 | 4,903 | 51.6 | 54 | 49.6 | 10.0 | 66.7 | 10.3 |
| 06 | 50 | 10,097 | 29.9 | 44 | 29.5 | 11.3 | 64.7 | 9.7 |
| 07 | 30 | 3,052 | 36.0 | 29 | 36.3 | 15.4 | 51.6 | 8.4 |
| 08 | 22 | 3,623 | 28.6 | 42 | 24.1 | 7.7 | 40.8 | 7.4 |
| 09 | 42 | 5,287 | 15.1 | 51 | 14.3 | 0.0 | 35.7 | 6.9 |
| 10 | 77 | 5,345 | 55.2 | 57 | 55.8 | 39.8 | 81.8 | 7.1 |
| 11 | 34 | 4,903 | 44.4 | 61 | 43.0 | 13.6 | 87.5 | 13.0 |
| 12 | 21 | 4,887 | 35.5 | 67 | 34.4 | 15.4 | 53.1 | 7.9 |
| 13 | 19 | 4,831 | 34.9 | 65 | 33.9 | 9.1 | 72.7 | 10.3 |
| 14 | 14 | 1,746 | 38.5 | 44 | 38.8 | 0.0 | 60.0 | 10.3 |
| Total | 370 | 58,224 | 35.8 | 82§ | 35.1 | 10.6 | 62.2 | 7.9 |

* 01=Mathematics and computer Science; 02=Physics; 03=Chemistry; 04=Earth sciences; 05=Biology; 06=Medicine; 07=Agricultural and veterinary sciences; 08=Civil Engineering; 09=Industrial and information engineering; 10=Ancient history, philology, literature and art history; 11=History, philosophy, pedagogy and psychology; 12=Law; 13=Economics and statistics; 14=Political and social sciences

** Scientific Disciplinary Sectors. The complete list is accessible at www.disp.uniroma2.it/laboratoriortt/testi/Indicators/ssd5.html, last accessed 22/10/2014.

† Data compilation excludes individual university UDA faculties with less than 10 staff members.

§ Data compilation excludes individual university UDA faculties with less than 30 staff members.

*Table 2: Incidence of female professors in the Italian academic system per academic rank and per UDA (in brackets % of total); data observed at 31/12/2011, elaborated from the MIUR database at , last accessed 22/10/2014*

| UDA | Assistant | | Associate | | Full | | Total |
|---|---|---|---|---|---|---|---|
| | Male | Female | Male | Female | Male | Female | |
| 01 | 788 | 535 (40.4) | 589 | 390 (39.8) | 763 | 170 (18.2) | 3,235 |
| 02 | 702 | 249 (26.2) | 637 | 148 (18.9) | 500 | 52 (9.4) | 2,288 |
| 03 | 583 | 781 (57.3) | 546 | 374 (40.7) | 527 | 130 (19.8) | 2,941 |
| 04 | 326 | 161 (33.1) | 247 | 111 (31) | 201 | 40 (16.6) | 1,086 |
| 05 | 887 | 1,542 (63.5) | 674 | 641 (48.7) | 813 | 346 (29.9) | 4,903 |
| 06 | 3,081 | 2,049 (39.9) | 2,154 | 694 (24.4) | 1,844 | 275 (13) | 10,097 |
| 07 | 738 | 656 (47.1) | 556 | 323 (36.7) | 659 | 120 (15.4) | 3,052 |
| 08 | 983 | 610 (38.3) | 825 | 271 (24.7) | 778 | 156 (16.7) | 3,623 |
| 09 | 1,758 | 465 (20.9) | 1,287 | 238 (15.6) | 1,442 | 97 (6.3) | 5,287 |
| 10 | 884 | 1,473 (62.5) | 692 | 884 (56.1) | 821 | 591 (41.9) | 5,345 |
| 11 | 1,004 | 1,102 (52.3) | 752 | 631 (45.6) | 972 | 442 (31.3) | 4,903 |
| 12 | 1,161 | 1,001 (46.3) | 729 | 408 (35.9) | 1,263 | 325 (20.5) | 4,887 |
| 13 | 1,057 | 891 (45.7) | 854 | 486 (36.3) | 1,234 | 309 (20) | 4,831 |
| 14 | 454 | 394 (46.5) | 305 | 170 (35.8) | 314 | 109 (25.8) | 1,746 |
| Total | 14,406 | 11,909 (45.3) | 10,847 | 5,769 (34.7) | 12,131 | 3,162 (20.7) | 58,224 |

## 3. Methods and data

We measure the research performance at the individual level by an indicator named Fractional Scientific Strength (FSS), which embeds both the number of publications produced and their standardized impact. To operationalize the measure of research performance we adopt several simplifications and assumptions. Because of a lack of data, we assume that the same resources are available to all professors in the same field. Because the intensity of publications varies across fields (Butler, 2007; Moed et al., 1985; Garfield, 1979), in order to avoid distortions in productivity rankings (Abramo et al., 2008) it is obligatory to compare researchers within the same field (SDS). In a previous study Abramo et al. (2011) demonstrated that productivity of full, associate and assistant professors is different. Because the distribution of gender by rank is not uniform, in order to avoid distortions we then need to compare professors' performance within the same academic rank. However given that in this study we intend to construct performance ranking lists by gender; if we accounted for the tripartite division by academic rank we would consider a situation in which we have a small number of female professors in a high number of SDSs, causing significance problems. Rather than accounting for the discrete divisions of academic rank we circumvent the problem through normalizing the performance by wage, the rationale being: the more one earns the more she/he is expected to produce[8].

In formula, the average yearly productivity of an individual, over a period of time,

---

[8] Wage effectively becomes a proxy for rank. In the Italian university system, salaries are in fact established at the national level and are fixed by academic rank and seniority. Thus all professors of the same academic rank and seniority receive the same wage, regardless of their merits and the university that employs them. The information on individual salaries is unavailable but the salaries ranges for rank and seniority are published. Thus we have approximated the wage for each individual as the national average of their academic rank.

accounting for the cost of labor, is[9]:

$$FSS = \frac{1}{w} * \frac{1}{t} \sum_{i=1}^{N} \frac{c_i}{\bar{c}} f_i$$

[1]

Where:
w = average yearly wage of the professor[10];
t = number of years of work of the professor in the period of observation;
N = number of publications of the professor in the period of observation;
$c_i$ = citations received by publication *i*;
$\bar{c}$ = average of the distribution of citations received for all cited publications[11] indexed in the same year and subject category of publication *i*;
$f_i$ = fractional contribution of the researcher to publication *i*.

We adopt the fractional counting method rather than full counting, as we believe it aligns with microeconomic theory of production. Moreover this methodology can account for the different contribution of authors where such is signaled by their positioning in the byline. Fractional contribution equals the inverse of the number of authors, in those fields where the practice is to place the authors in simple alphabetical order, but assumes different weights in other cases. For the life sciences, widespread practice in Italy and abroad is for the authors to indicate the various contributions to the published research by the order of the names in the byline. For these disciplines, we give different weights to each co-author according to their order in the byline and the character of the co-authorship (intra-mural or extra-mural). If first and last authors belong to the same university, 40% of citations are attributed to each of them; the remaining 20% are divided among all other authors. If the first two and last two authors belong to different institutions, 30% of citations are attributed to first and last authors; 15% of citations are attributed to second and last author but one; the remaining 10% are divided among all others[12]. Failure to account for the number and position of authors in the byline would result in notable ranking distortions at the individual level (Abramo et al. 2013b).

We calculate $FSS_R$ of each professor in each SDS and express it on a percentile scale of 0-100 (worst to best) for comparison with the performance of all Italian colleagues of the same SDS; or as the ratio to the average productivity of all Italian colleagues of the same SDS with productivity above zero[13]. To ensure the reliability of the results issuing from the evaluation, we have chosen a long enough observation period: 2006-2010 (Abramo et al., 2012b). Given the characteristics of the Italian university system, we can safely exclude that the productivity ranking lists may be distorted by variable returns to scale, due to different sizes of universities (Abramo et al. 2012c) or by returns to scope of research fields (Abramo et al. 2013c).

---

[9] A thorough description of the formula, underlying theory, assumptions and limits may be found in Abramo and D'Angelo (2014).
[10] We assume that other production factors are equally available to all researchers. If not, their value should be taken into account.
[11] A preceding article by the same authors demonstrated that the average of the distribution of citations received for all cited publications of the same year and subject category is the most effective scaling factor (Abramo et al. 2012a).
[12] The weighting values were assigned following advice from senior Italian professors in the life sciences. The values could be changed to suit different practices in other national contexts.
[13] In a preceding article the authors demonstrated that the average of the productivity distribution of researchers with productivity above 0 is the most effective scaling factor to compare the performance of researchers of different fields (Abramo et al. 2013d).

Data on research staff of each university, such as years of employment in the observed period, academic rank and their SDS classification are extracted from the database on Italian university personnel, maintained by the Ministry for Universities and Research[14]. Unfortunately, information on leaves of absence is not available and cannot be accounted for in the calculation of yearly productivity, to the disadvantage of women on maternity leave in the period of observation.

The bibliometric dataset for the analysis draws on the Observatory of Public Research (ORP), a database developed and maintained by the authors and derived under license from the WoS. Beginning from the raw data of Italian publications indexed in WoS, and applying a complex algorithm for disambiguation of the true identity of the authors and their institutional affiliations (for details see D'Angelo et al., 2011), each publication is attributed to the university professor that produced it, with a harmonic average of precision and recall (F-measure) equal to 96 (error of 4%). Beginning from this data we are able to calculate FSS for each Italian professor. For the WoS-indexed publications to serve as a more robust proxy of overall output of a researcher, the field of observation is limited to those SDSs (188 in all) where at least 50% of member scientists produced at least one publication in the period 2006-2010[15]. For the purposes of the study and to ensure significant representation of both genders in each field, we then further limit the analysis to those SDSs (99 in all) with at least 30 individuals of each gender. Table 3 shows the final dataset.

*Table 3: Dataset for the analysis: number of fields (SDSs), universities and professors in each UDA under investigation*

| UDA | N. of SDSs | Universities | Professors* | Of which female |
|---|---|---|---|---|
| Mathematics and computer science | 8 | 65 | 3,297 | 1,105 (33.5) |
| Physics | 4 | 61 | 2,161 | 390 (18.0) |
| Chemistry | 9 | 59 | 3,199 | 1,212 (37.9) |
| Earth sciences | 4 | 41 | 534 | 176 (33.0) |
| Biology | 19 | 66 | 5,338 | 2,591 (48.5) |
| Medicine | 29 | 60 | 9,426 | 2,805 (29.8) |
| Agricultural and veterinary sciences | 17 | 43 | 2,163 | 755 (34.9) |
| Civil engineering | 3 | 49 | 828 | 130 (15.7) |
| Industrial and information engineering | 6 | 64 | 2,051 | 298 (14.5) |
| Total | 99 | 79 | 28,997 | 9,462 (32.6) |

\* *The number of professors may be higher than observed in Table 1. Table 1 refers to the stock of the Italian academic population as of the moment 31/12/2011, whereas the dataset includes professors with at least three years of seniority over the period 2006-2010.*

## 4. Analysis and discussion

In this section we compare the distributions of research performance for males and females at the SDS and UDA levels. Then for each SDS we construct a first ranking list by FSS, for all professors without gender distinction.

---

[14] http://cercauniversita.cineca.it/php5/docenti/cerca.php. Last accessed 22/10/2014.
[15] The complete list is accessible at http://www.disp.uniroma2.it/laboratoriortt/testi/Indicators/ssd5.html, last accessed 22/10/2014.

We then measure the FSS of each female professor and normalize it by the mean of all productive female professors, also doing the same for males. Proceeding from these steps we construct a second "altogether" ranking list by FSS normalized by the respective means, where the manner of building the list accounts for gender differences.

Finally we compare the two ranking lists: with and without gender differences.

**4.1 Gender differences in research productivity**

As an example, in Figure 1 we represent the distribution of FSS by gender in Biology. The graph does not consider researchers with zero values in FSS. To obtain a more symmetrical distribution and not the classic rightly skewed view of scientific performance, we considered the logs of FSS values. To represent the distribution we use the Epanechnikov kernel function, a non-parametric means of estimating the probability density function of a continuous variable. We observe that the right tail of the distribution is dominated by males, while the mode values (represented by peaks of distributions) are quite similar, as are the forms of the distributions themselves.

*Figure 1: FSS distribution by gender in Biology*

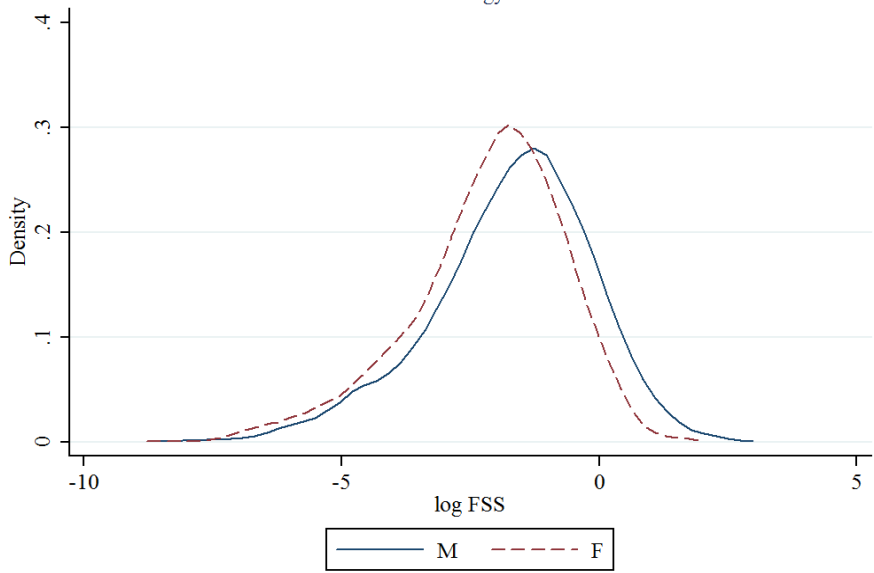

Table 4 shows a deeper investigating in the UDA, providing the descriptive statistics of the distribution of FSS for all researchers within each SDS (including those with no production). We test the difference between the genders in terms of unproductives and the median value of FSS. For the first case we use the classic z-test for proportions; in the second case we use Student's t-test for independent samples. No significant difference is seen the percentages of unproductive researchers by gender except in BIO/15 (p-value<0.10). In roughly half the SDSs (8 of 19) the median value of FSS is higher for males, with a high level of significance (p-value < 0.01); in a further two cases the significance is weaker. In the remaining nine SDSs there are no significant gender differences. There are only two SDSs, BIO/01 and BIO/02, where women faculty register a higher productivity than men, although the difference is still not significant. In these SDSs the greater productivity of women is further indicated by the median value of the distribution, and by the maximum values observed (4.715 women vs. 2.004 men in BIO/01 and 2.866 vs. 1.042 in BIO/02). These two SDSs are also unusual in their dispersions, being the only ones where the standard deviation of female performance is greater than for males.

*Table 4: Descriptive statistics of FSS in Biology by gender*

| SDS‡ | Obs | Unproductive % F | Unproductive % M | Mean F | Mean M | Median F | Median M | Max F | Max M | Std. dev. F | Std. dev. M |
|---|---|---|---|---|---|---|---|---|---|---|---|
| BIO/01 | 137 | 18.1 | 18.5 | 0.271 | 0.252 | 0.122 | 0.119 | 4.715 | 2.004 | 0.602 | 0.380 |
| BIO/02 | 134 | 35.8 | 32.8 | 0.163 | 0.101 | 0.038 | 0.019 | 2.866 | 1.042 | 0.418 | 0.220 |
| BIO/03 | 145 | 24.6 | 33.8 | 0.123 | 0.207 | 0.043 | 0.034 | 0.727 | 2.514 | 0.173 | 0.433 |
| BIO/04 | 113 | 9.6 | 3.3 | 0.244 | 0.416 | 0.189 | 0.165 | 1.344 | 3.551 | 0.251 | 0.643 |
| BIO/05 | 323 | 14.2 | 15.8 | 0.237 | 0.374** | 0.111 | 0.211 | 4.333 | 3.744 | 0.462 | 0.475 |
| BIO/06 | 260 | 12.8 | 11.6 | 0.192 | 0.361** | 0.092 | 0.116 | 3.718 | 5.581 | 0.385 | 0.708 |
| BIO/07 | 240 | 15.6 | 14.7 | 0.255 | 0.431*** | 0.156 | 0.194 | 1.906 | 4.467 | 0.346 | 0.682 |
| BIO/08 | 77 | 28.1 | 35.6 | 0.085 | 0.107 | 0.023 | 0.035 | 0.418 | 0.540 | 0.119 | 0.139 |
| BIO/09 | 647 | 11.1 | 9.8 | 0.233 | 0.405*** | 0.118 | 0.165 | 2.484 | 6.320 | 0.331 | 0.707 |
| BIO/10 | 924 | 7.1 | 5.4 | 0.231 | 0.430*** | 0.111 | 0.186 | 3.817 | 9.228 | 0.362 | 0.850 |
| BIO/11 | 221 | 7.0 | 5.8 | 0.247 | 0.614*** | 0.095 | 0.236 | 2.197 | 7.361 | 0.356 | 1.078 |
| BIO/12 | 164 | 1.4 | 5.3 | 0.211 | 0.841*** | 0.122 | 0.221 | 1.724 | 14.864 | 0.282 | 2.118 |
| BIO/13 | 272 | 7.7 | 11.5 | 0.245 | 0.343 | 0.111 | 0.124 | 1.709 | 3.251 | 0.328 | 0.571 |
| BIO/14 | 699 | 6.9 | 6.2 | 0.367 | 0.595*** | 0.197 | 0.295 | 5.469 | 9.246 | 0.604 | 0.947 |
| BIO/15 | 95 | 9.1 | 0.0 | 0.380 | 0.572 | 0.222 | 0.175 | 1.610 | 5.328 | 0.424 | 0.945 |
| BIO/16 | 371 | 10.0 | 12.6 | 0.203 | 0.422*** | 0.087 | 0.132 | 2.821 | 8.592 | 0.349 | 0.906 |
| BIO/17 | 193 | 4.5 | 7.2 | 0.208 | 0.595*** | 0.108 | 0.143 | 2.440 | 6.840 | 0.338 | 1.288 |
| BIO/18 | 214 | 6.5 | 3.8 | 0.220 | 0.316 | 0.119 | 0.176 | 1.802 | 2.918 | 0.287 | 0.456 |
| BIO/19 | 109 | 1.7 | 4.1 | 0.255 | 0.318 | 0.151 | 0.173 | 1.499 | 1.865 | 0.290 | 0.364 |

‡ *Complete list of acronyms at www.disp.uniroma2.it/laboratoriortt/testi/Indicators/ssd5.html, last accessed 22/10/2014*
*\*\* p-value<0.05;\*\*\* p-value<0.01*

We can readily observe that the distributions of FSS are remarkably different between the SDSs, due to the different intensity of publication and citation typical of the different fields of research.

We extend the analysis above to the remaining eight UDAs. Figure 2 presents the distributions of FSS by gender, graphed excluding researchers with zero value of FSS. We observe a certain similarity between the distributions and in all areas, except for Medicine and Civil engineering, where the tails of outliers are completely dominated by males. The average and median values (net of unproductives) are slightly in favor of the male gender.

Including researchers with an FSS value of zero, we now calculate the descriptive statistics for the nine UDAs (Table 5). We test the difference between genders for percentage of unproductive professors and mean value of FSS, respectively using the z-test and Student t-test, as previously. The gender differences for incidence of unproductive professors are significant in only two UDAs: in Agricultural and veterinary sciences, where the incidence of unproductive professors is higher for males; in Mathematics and computer sciences, where the opposite holds. Concerning gender differences between the mean values of FSS, these are significant in all UDAs except Industrial and information engineering, and always in favor of men.

*Figure 2: FSS distribution by gender in eight UDAs*

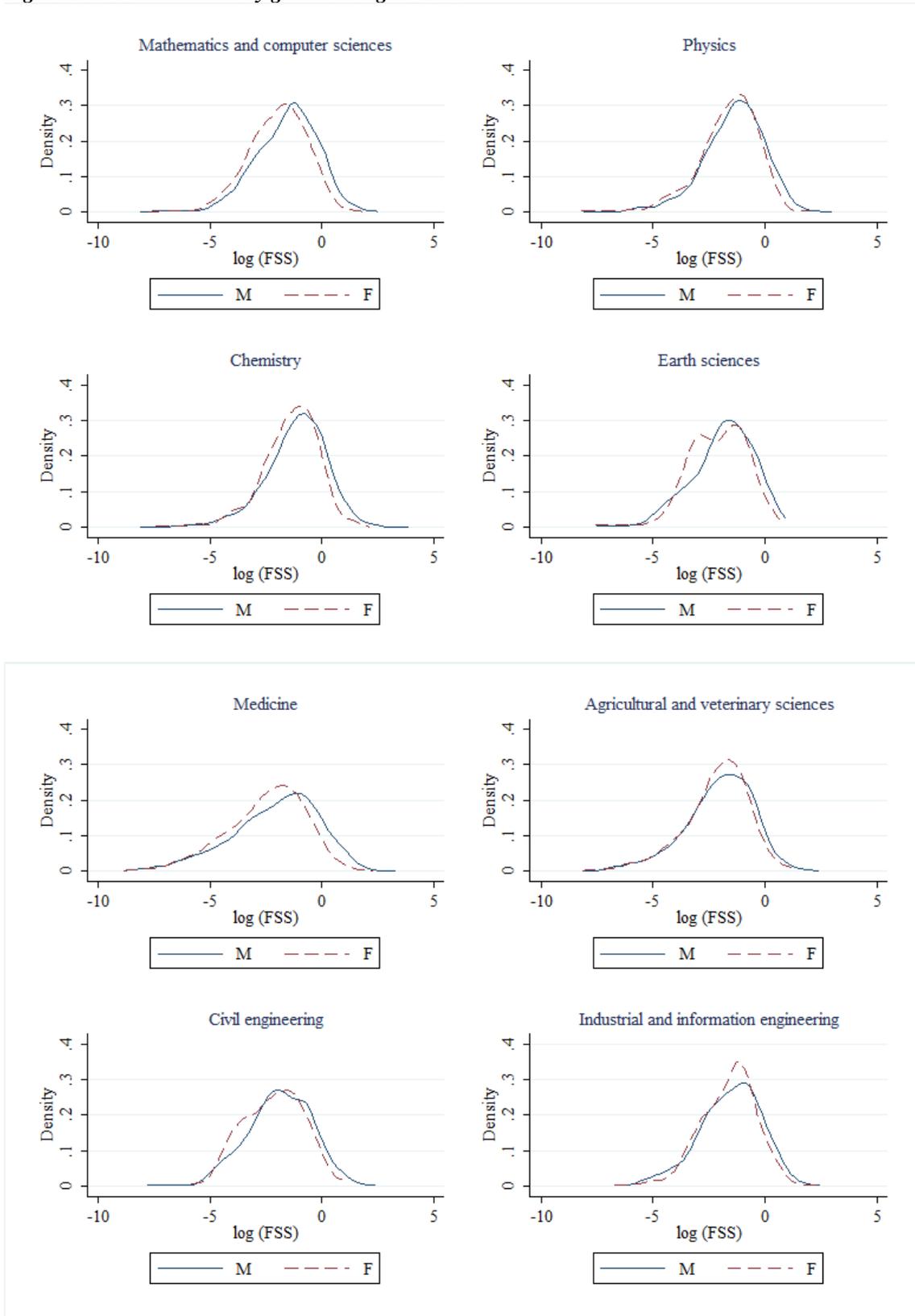

*Table 5: Descriptive statistics of FSS by gender in each UDA*

| UDAS† | Gender | % of non productive | Mean | Median | IQR | Standard deviation |
|---|---|---|---|---|---|---|
| 01 | M | 24.0*** | 0.351*** | 0.133 | 0.406 | 0.647 |
|    | F | 30.0 | 0.197 | 0.067 | 0.243 | 0.357 |
| 02 | M | 9.3 | 0.489*** | 0.246 | 0.515 | 0.801 |
|    | F | 9.7 | 0.339 | 0.205 | 0.394 | 0.45 |
| 03 | M | 4.8 | 0.672*** | 0.342 | 0.658 | 1.309 |
|    | F | 4.1 | 0.445 | 0.268 | 0.47 | 0.593 |
| 04 | M | 14.8 | 0.267** | 0.142 | 0.35 | 0.351 |
|    | F | 15.9 | 0.201 | 0.09 | 0.255 | 0.275 |
| 05 | M | 10.5 | 0.435*** | 0.171 | 0.426 | 0.863 |
|    | F | 10.1 | 0.246 | 0.117 | 0.266 | 0.405 |
| 06 | M | 18.2 | 0.392*** | 0.091 | 0.392 | 0.851 |
|    | F | 17.5 | 0.213 | 0.06 | 0.228 | 0.445 |
| 07 | M | 15.4** | 0.275*** | 0.111 | 0.313 | 0.524 |
|    | F | 11.5 | 0.219 | 0.114 | 0.253 | 0.324 |
| 08 | M | 35.2 | 0.237** | 0.055 | 0.251 | 0.54 |
|    | F | 30.8 | 0.161 | 0.051 | 0.21 | 0.264 |
| 09 | M | 17.7 | 0.388 | 0.165 | 0.444 | 0.652 |
|    | F | 14.4 | 0.336 | 0.181 | 0.352 | 0.614 |
| Total | M | 15.9*** | 0.414*** | 0.15 | 0.444 | 0.843 |
|       | F | 14.5 | 0.258 | 0.112 | 0.293 | 0.448 |

*† 01=Mathematics and computer Science; 02=Physics; 03=Chemistry; 04=Earth sciences; 05=Biology; 06=Medicine; 07=Agricultural and veterinary sciences; 08=Civil Engineering; 09=Industrial and information engineering; 10=Ancient history, philology, literature and art history; 11=History, philosophy, pedagogy and psychology; 12=Law; 13=Economics and statistics; 14=Political and social sciences*

*\*\* p-value<0.05; \*\*\* p-value<0.01*

## 4.2 Correlation between gender and performance

The preliminary analysis shows important differences between genders in scientific performance of the researchers. To estimate the magnitude of association between gender and scientific performance (FSS) we apply a point-biserial coefficient (Tate, 1954), which permits analysis of the relationship between a continuous variable (FSS) and a binary variable (gender). We apply this coefficient at two levels: SDSs and UDAs. For each level, the coefficient of correlation ($r_{pb}$) between FSS and gender is expressed:

$$r_{pb} = \frac{\overline{FSS}_M - \overline{FSS}_F}{SD} * \sqrt{\frac{n_M * n_F}{N(N-1)}}$$

[2]

where:
$\overline{FSS}_M$: average value of FSS of male individuals.
$\overline{FSS}_F$: average value of FSS of female individuals.
SD: standard deviations of entire distribution of FSS.
$n_M$: number of males.
$n_M$: number of females.

Verification of the null hypothesis that the coefficient is equal to zero is performed by a procedure similar to Pearson's correlation.

In Table 6 we observe that the values of correlations at the UDA level are significant in seven cases, although with variation among UDAs and never at very high intensity.

Biology and Mathematics show the highest correlations, respectively at 0.149 and 0.133; Agricultural and veterinary sciences and Physics show the lowest correlations.

At the lower level of the SDSs, there is no disciplinary area where the SDSs show a link between gender and performance. In Civil engineering in particular, none of the 3 SDSs analyzed shows a value of correlation at 5% significance.

*Table 6: Descriptive statistics of correlation coefficients*

| UDA | Analysis at second level (SDS) | | | Analysis at first level (UDA) |
|---|---|---|---|---|
| | % of significant SDSs** | Min($r_{pb}$) | Max($r_{pb}$) | $r_{pb}$ |
| Mathematics and computer science | 14.3 | 0.039 | 0.248 | 0.133** |
| Physics | 2.4 | 0.036 | 0.106 | 0.082** |
| Chemistry | 32.1 | 0.027 | 0.206 | 0.105** |
| Earth sciences | 10.7 | 0.032 | 0.216 | 0.104** |
| Biology | 42.9 | -0.130 | 0.227 | 0.149** |
| Medicine | 67.9 | -0.027 | 0.270 | 0.123** |
| Agricultural and veterinary sciences | 21.4 | -0.001 | 0.216 | 0.076** |
| Civil engineering | 0.0 | 0.071 | 0.153 | 0.084 |
| Industrial and information engineering | 8.3 | -0.030 | 0.118 | 0.041 |
| Total | 54.5 | -0.130 | 0.270 | 0.113** |

*\*\* p-value<0.05; \*\*\* p-value<0.01*

### 4.3 Accounting for gender in research performance ranking lists

In this section we measure the effect of distinction by gender on the ranking of individuals for their bibliometric performance. Our specific question is: how does the position of a scientist in ranking change if his or her performance is compared to the performances of others in the same gender, in respect to a ranking with no distinction by gender. To reach the objective we calculate the ratio of the value of performance indicator for a scientist (FSS) to the mean of the SDS distribution. As an example, let us assume that in a given SDS: there are 50 female and 70 male professors; the mean FSS of the overall population is 2; while the mean FSS of the female subpopulation is 1.8, and the mean FSS of the male subpopulation is 2.2. The FSS of a certain female professor is 2 and the FSS of a certain male professor is 2.2. In an overall ranking list undistinguished for gender the male professor would rank higher than the female, because his FSS is 10% higher than hers. Differently, if the ranking list distinguishes by gender it is the female professor who would rank higher: her distance from the female mean FSS is above 10%, while the male professor FSS is exactly at the mean of the male population. We first provide a graphic representation of the differences in rank between the two methods of comparative performance evaluation, then follow with a quantitative analysis.

### 4.3.1. Visualization of rank differences

Figure 3 presents the example of the analysis of the distances from mean FSS for the Biology UDA. Each point represents one individual: red circles for female professors; blue crosses for males. For each individual we show two items of information: along the *x* axis is the ratio of the individual's value for FSS to the mean FSS for all professors of their SDS, without distinction for gender; on the *y* axis is the ratio between their FSS value to the mean for their own gender in the SDS. If all of these values lay along the graph bisector this would indicate that distinction by gender had no effect on the individual ranks. When a value appears above the bisector, this indicates that the individual receives an improvement in rank when performance is distinguished by gender, compared to under evaluation without distinction for gender. We can clearly observe that 99% of all individuals who place above the bisector are women: these are the individuals who benefit from an evaluation by gender.

*Figure 3: Distances from gender and total mean FSS in Biology (4,787 observations)*

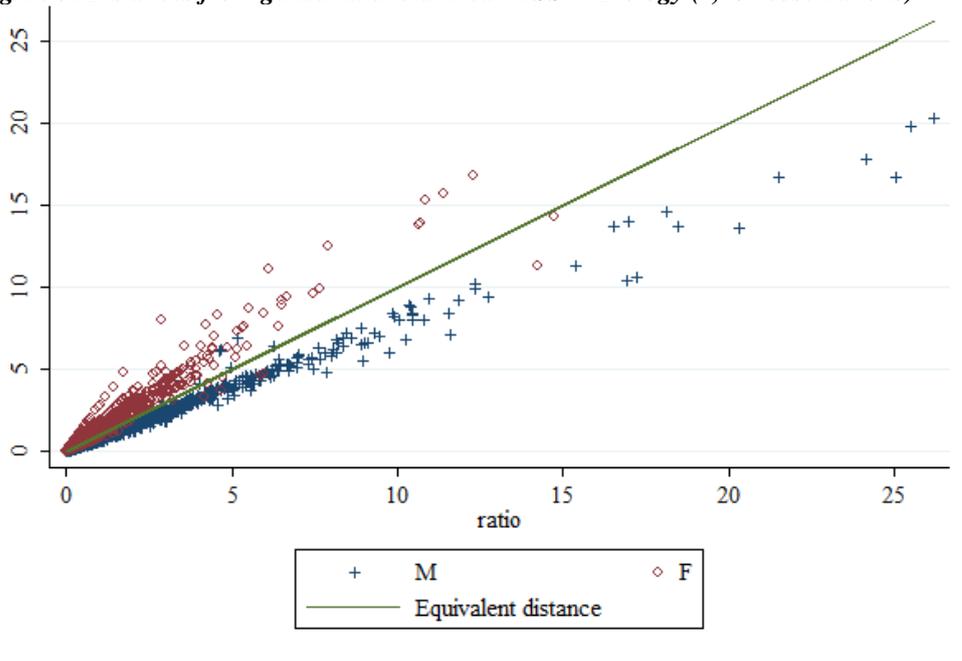

We extend the graphic analysis to the remaining eight UDAs, as seen in Figure 5, obtaining an overall view. We can observe that in the disciplines of Agricultural and veterinary sciences, Mathematics, Earth sciences and Medicine, female researchers obtain particular benefit from a ranking by gender. In the remaining UDAs benefits are lower. For example, in Physics there is an outlier that increases her distance from the mean by roughly 60% when compared to the mean for female gender. In contrast, in Agricultural and veterinary science there are two male outliers who experience a 30% drop in distance when their performance is compared to the mean of their gender.

*Figure 4: Distances from gender and total mean FSS in eight UDAs*

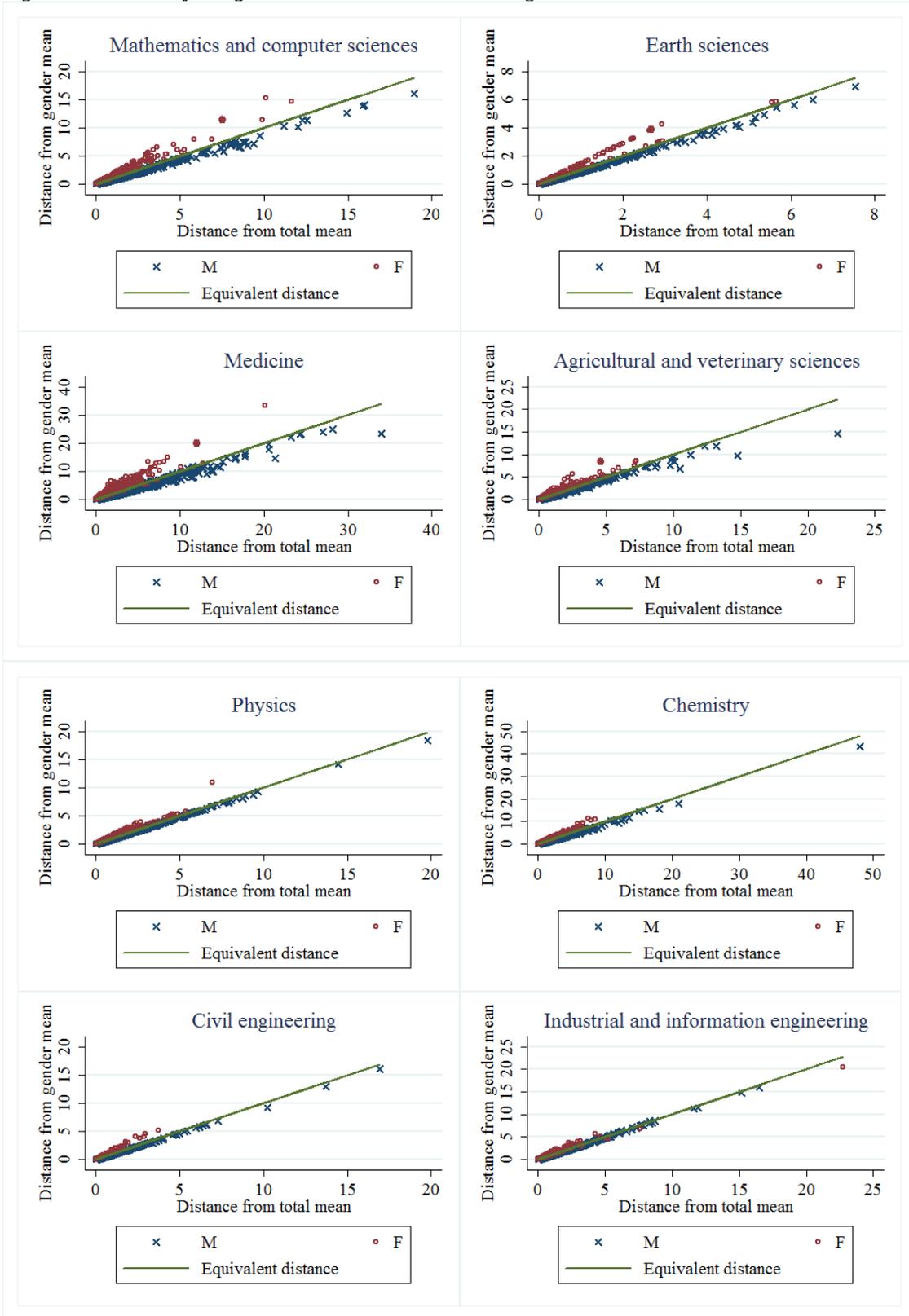

**4.3.2 Quantification of rank differences**

For each SDS we construct two ranking lists by FSS distances from the mean, with and without distinction by gender, expressing the rank of each individual in percentiles (100 = top). For each individual we then calculate the difference between their positions under the two ranking lists. Table 7 presents the example of the descriptive statistics synthesizing the variations in rank of each individual in the Biology UDA. The data bring out that in 17 out of 19 SDSs, the female gender on average gains positions, in the ranking list distinguished by gender. In detail, each female researcher gains an average of 2.5 percentile positions in BIO7/13 and almost 13 positions in BIO/12. The SDSs of BIO/01 and BIO/02 are the only cases where female researchers show on average a disadvantage (although very contained in BIO/01). In contrast, male researchers lose between 3.3 percentile positions in BIO/15 and almost 11 positions in BIO/17.

*Table 7: Descriptive statistics of percentile rank shifts in the 19 SDSs of Biology*

| SDS‡ | Male | | | | | Female | | | | |
|---|---|---|---|---|---|---|---|---|---|---|
| | Mean | Median | St.dev. | Min | Max | Mean | Median | St.dev. | Min | Max |
| BIO/01 | 0.747 | 0.9 | 0.764 | 0 | 3 | -0.671 | -0.9 | 0.812 | -4 | 0 |
| BIO/02 | 4.553 | 3.4 | 3.564 | 0 | 14 | -4.844 | -5.7 | 2.253 | -9 | 0 |
| BIO/03 | -5.928 | -6.9 | 3.047 | -10 | 0 | 6.404 | 6.9 | 2.535 | 0 | 12 |
| BIO/04 | -4.663 | -4.8 | 3.524 | -13 | 0 | 5.806 | 5.7 | 2.768 | 0 | 11 |
| BIO/05 | -3.761 | -4.0 | 1.881 | -7 | 0 | 6.230 | 6.2 | 3.108 | 0 | 12 |
| BIO/06 | -7.127 | -7.9 | 3.796 | -14 | 0 | 5.447 | 5.7 | 2.375 | 0 | 10 |
| BIO/07 | -4.376 | -3.65 | 2.815 | -11 | 0 | 7.363 | 7.4 | 2.978 | 1 | 12 |
| BIO/08 | -3.455 | -3.9 | 1.703 | -6 | 0 | 4.357 | 3.9 | 3.334 | 0 | 12 |
| BIO/09 | -4.468 | -4.5 | 2.293 | -8 | 0 | 6.314 | 6.9 | 2.418 | 0 | 10 |
| BIO/10 | -6.002 | -6.6 | 2.808 | -11 | 0 | 6.088 | 6.9 | 2.998 | 0 | 10 |
| BIO/11 | -7.863 | -8.7 | 3.292 | -13 | 0 | 9.604 | 11.7 | 4.203 | 1 | 15 |
| BIO/12 | -10.080 | -10.2 | 5.166 | -20 | 0 | 12.981 | 14.6 | 5.361 | 3 | 20 |
| BIO/13 | -4.351 | -4.9 | 2.005 | -8 | 0 | 2.564 | 2.4 | 1.279 | 0 | 6 |
| BIO/14 | -5.096 | -6.0 | 2.625 | -10 | 0 | 4.768 | 4.4 | 2.531 | 0 | 9 |
| BIO/15 | -3.365 | -3.4 | 2.519 | -10 | 0 | 2.688 | 2.2 | 1.569 | 0 | 6 |
| BIO/16 | -6.398 | -6.4 | 3.030 | -11 | 0 | 6.566 | 6.25 | 3.195 | 0 | 12 |
| BIO/17 | -10.938 | -11.0 | 6.422 | -21 | -1 | 8.000 | 8.8 | 2.850 | 0 | 12 |
| BIO/18 | -3.502 | -4.0 | 1.883 | -8 | 0 | 3.504 | 3.5 | 1.707 | 0 | 6 |
| BIO/19 | -3.421 | -2.9 | 1.812 | -8 | 0 | 2.719 | 2.9 | 1.608 | 0 | 7 |

‡ *Complete list of acronyms at www.disp.uniroma2.it/laboratoriortt/testi/Indicators/ssd5.html, last accessed 22/10/2014*

These analyses were repeated for all 99 SDSs under examination. Focusing attention on the average number of positions gained or lost for each gender, in Table 8 we show the division of the SDSs in classes constructed for average intensity of variation. Classes 1 to 3 contain SDSs with negative average variation; classes 4 to 6 contain SDSs with positive variation. In total, in 93 out of 99 SDSs, the female staff obtains an average shift up under the ranking lists with distinction by gender (the count of all SDSs in classes 4 to 6). The indications that emerge are as clear as in the preceding graphic analysis, with some further details. In Industrial and information engineering, there are two SDSs with average negative variation for female researchers, alongside four SDS with positive variation; this discipline and the Biology UDA also result as the only cases where there is more than one SDS with negative variation in ranking for women. Medicine shows the highest total number of SDSs (12) with average positive variation greater than 8 percentile points. For the male gender, the only SDS that shows an average positive variation of more than 8 percentiles is found in Medicine, while a full 6 SDSs, scattered in Biology (2), Medicine (3) and Agricultural and veterinary sciences (1), present an average negative variation of over 8 percentiles. In total, there are only seven cases in the 99 SDSs examined where men gain any benefit from comparative evaluation distinguished by gender.

*Table 8: Number of SDSs divided by class for average variation in percentile ranking*

| | Male | | | | | | Female | | | | | |
|---|---|---|---|---|---|---|---|---|---|---|---|---|
| | - | | | + | | | - | | | + | | |
| UDAS† | Cl-1* | Cl-2 | Cl-3 | Cl-4 | Cl-5 | Cl-6 | Cl-1* | Cl-2 | Cl-3 | Cl-4 | Cl-5 | Cl-6 |
| 01 | | 3 | 5 | | | | | | | 2 | 2 | 4 |
| 02 | | | 4 | | | | | | | 2 | | 2 |
| 03 | | 5 | 4 | | | | | | | 4 | 4 | 1 |
| 04 | | 1 | 3 | | | | | | | 1 | 2 | 1 |
| 05 | 2 | 10 | 5 | 1 | 1 | | | 1 | 1 | 4 | 10 | 3 |
| 06 | 3 | 7 | 17 | 1 | | 1 | | | 1 | 3 | 13 | 12 |
| 07 | 1 | 2 | 13 | 1 | | | | | 1 | 7 | 6 | 3 |
| 08 | | | 3 | | | | | | | | 1 | 2 |
| 09 | | | 4 | 2 | | | | | 2 | 1 | 2 | 1 |
| Total | 6 | 28 | 58 | 5 | 1 | 1 | | 1 | 5 | 24 | 40 | 29 |

† 01=Mathematics and computer Science; 02=Physics; 03=Chemistry; 04=Earth sciences; 05=Biology; 06=Medicine; 07=Agricultural and veterinary sciences; 08=Civil Engineerin; 09=Industrial and information engineering; 10=Ancient history, philology, literature and art history; 11=History, philosophy, pedagogy and psychology; 12=Law; 13=Economics and statistics; 14=Political and social sciences

\*     *Class-1: < -8*
      *Cl-2: ≥-8 and <-4*
      *Cl-3: ≥-4 and <0*
      *Cl-4: ≥ 0 and <+4*
      *Cl-5: ≥ +4 and <+8*
      *Cl-6: ≥ +8*

In conclusion, Table 9 provides the descriptive statistics per UDA, as previously developed in Table 7 for the case of the Biology UDA. In all disciplinary areas the female gender obtains on average a better position: the female researchers in the Civil engineering area gain an average of 8.4 positions, while women in Physics gain 4.5 positions. The male counterparts that lose the greatest number of positions under gender-distinguished ranking are the professors in Biology (-5.2 positions), while those in Industrial and information engineering show the least shift (average -0.4 positions).

In a full five out of the nine UDAs (01 to 04, 08) no male professor gains positions in moving from "altogether" ranking to ranking distinguished by gender. In contrast, the analysis registers "maximum gains" for female professors that oscillate between 14.3 in Earth science and 28.9 in Medicine.

*Table 9: Descriptive statistics of percentile ranking shifts in the 9 UDAs*

|  | Male | | | | | Female | | | | |
| --- | --- | --- | --- | --- | --- | --- | --- | --- | --- | --- |
| UDAS† | Mean | Median | St.dev. | Min | Max | Mean | Median | St.dev. | Min | Max |
| 01 | -3.39 | -2.9 | 2.60 | -15.0 | 0.0 | 7.27 | 6.9 | 3.86 | 0.0 | 18.4 |
| 02 | -1.00 | -0.8 | 0.90 | -3.7 | 0.0 | 4.54 | 3.4 | 4.18 | 0.0 | 17.0 |
| 03 | -3.24 | -2.2 | 2.80 | -17.9 | 0.0 | 5.26 | 5.0 | 3.34 | 0.0 | 17.9 |
| 04 | -2.56 | -2.0 | 2.17 | -8.8 | 0.0 | 5.24 | 5.3 | 4.23 | 0.0 | 14.3 |
| 05 | -5.23 | -5.3 | 3.80 | -21.0 | 13.8 | 5.50 | 5.4 | 3.84 | -9.2 | 19.7 |
| 06 | -3.04 | -2.2 | 2.97 | -21.1 | 2.3 | 7.09 | 6.4 | 4.88 | -4.6 | 28.9 |
| 07 | -2.61 | -1.6 | 3.42 | -20.6 | 1.1 | 4.63 | 3.4 | 4.73 | -1.1 | 24.5 |
| 08 | -1.71 | -1.5 | 1.33 | -8.2 | 0.0 | 8.41 | 8.4 | 3.31 | 2.0 | 17.6 |
| 09 | -0.42 | -0.3 | 0.88 | -3.4 | 2.1 | 2.40 | 1.1 | 4.94 | -5.2 | 17.0 |

† *01=Mathematics and computer Science; 02=Physics; 03=Chemistry; 04=Earth sciences; 05=Biology; 06=Medicine; 07=Agricultural and veterinary sciences; 08=Civil Engineering; 09=Industrial and information engineering; 10=Ancient history, philology, literature and art history; 11=History, philosophy, pedagogy and psychology; 12=Law; 13=Economics and statistics; 14=Political and social sciences*

## 5. Conclusions

The data on the Italian context show that that the presence of women in the national academic staff is limited to 35.8%, and is not homogenous across scientific fields, being particularly low in the hard sciences. The gender distribution by academic rank reveals a remarkable trend towards decrease in female underrepresentation: the incidence of women within assistant-professor rank is much higher than it is for full professors. Measuring the research performance of professors in the hard sciences by a bibliometric indicator that embeds both number of publications and their impact (FSS), we observe noticeable gender differences between average values of performance (higher for men). There are no notable differences in terms of rates for unproductive professors.

The study inquired into the issue of gender in procedures for comparative evaluation among single researchers. In particular it analyzed the variations of rank between two lists: one that did not distinguish by gender and another that did. For each research field we constructed two ranking lists by FSS distances from the mean (i.e. lists with and without distinction by gender) and expressed the ranks of the individual in percentiles. For each individual we then calculated the difference in positions between the two ranking lists.

What emerges is a heterogeneous panorama, even within the same discipline (UDA). However in roughly 70% of the individual disciplines (SDSs) analyzed, women professors obtain a shift ahead in the ranking lists with distinction by gender. The disciplines with the most substantial average shifts are: Civil engineering, Mathematics, Medicine.

The aim of the current paper was not to further investigate if or to what extent there is gender discrimination in the research sphere, or to further examine the objective limitations on women's careers given their social roles. Nor do we intend to issue *a priori* recommendations on the suitability of conducting comparative evaluation research performance that would take account of gender. Our current objective was to clarify whether the comparison of individuals' research performance between peers of the same gender leads to rank positions that are detectably different compared to those from ranking lists constructed without distinction by gender. We leave it to the decision-maker to choose which approach to adopt, given the objectives of the evaluation and the conditions of the context. Knowing the extent of position differences between the two ranking approaches may help the management make more informed decisions. Possible future directions of research concern the verification of the differences in rank in the evaluation of organizational units, such as research groups, departments and research institutions.